\documentstyle[12pt, preprint]{aastex}
\def\half{\textstyle {1 \over 2} \displaystyle}

\def\ltsima{$\;\buildrel < \over \sim \;$}
\def\simlt{\lower.5ex \hbox{\ltsima}}
\def\gtsima{$\;\buildrel > \over \sim \;$}
\def\simgt{\lower.5ex \hbox{\gtsima}}
\def\kms{{\rm \, km \, s^{-1}}}

\begin{document}
\title{Submillimeter Wave Astronomy Satellite mapping observations of
water vapor around Sagittarius B2}

\author{David A. Neufeld\altaffilmark{1}, Edwin A. Bergin\altaffilmark{2},
Gary J. Melnick\altaffilmark{2}  and 
Paul F. Goldsmith\altaffilmark{3}}

  
\altaffiltext{1}{Department of Physics \& Astronomy,  The Johns Hopkins University, 3400
North Charles Street,  Baltimore, MD 21218}

\altaffiltext{2}{Harvard-Smithsonian Center for Astrophysics, 60 Garden Street,
Cambridge, MA 02138}

\altaffiltext{3}{NAIC, Department of Astronomy, Cornell
University, Ithaca, NY 14853}



\begin{abstract}

Observations of
the $1_{10} - 1_{01}$ 556.936~GHz
transition of ortho-water
with the {\it Submillimeter Wave
Astronomy Satellite} (SWAS) have revealed
the presence of widespread emission and 
absorption by water vapor around the strong 
submillimeter continuum source Sagittarius B2. 
An incompletely-sampled spectral line map of a region 
of size $26 \times 19$ arcmin around Sgr B2 
reveals three noteworthy features.
First,  
absorption by foreground water vapor is
detectable at local standard-of-rest (LSR) 
velocities in the range --100 to 0 km/s at almost
every observed position.
Second, spatially-extended emission by water is detectable
at LSR velocities in the range 80 to 120 km/s
at almost every observed position.
This emission is attributable to the
180-pc molecular ring identified from previous
observations of CO.  The typical peak antenna 
temperature of 0.075~K for this component
implies a typical water abundance of 
1.2 -- 8 $\times 10^{-6}$ relative
to H$_2$. 
Third, strong absorption by water is observed 
within 5 arcmin of Sgr B2 at LSR velocities
in the range 60 to 82 km/s.  An analysis
of this absorption yields a
H$_2^{16}$O abundance $\sim 2 - 4 \times 10^{-7}$ relative
to H$_2$
if the absorbing water vapor is located within the
core of Sgr B2 itself; or, alternatively,
a water column density of $\sim 2.5 - 4 \times 10^{16}\,\rm
cm^{-2}$ if the water absorption originates in
the warm, foreground layer of gas proposed previously 
as the origin of ammonia absorption observed
toward Sgr B2.

\end{abstract}

\section{Introduction}

The Sagittarius B2 cloud represents the largest condensation
of molecular gas and dust in the Galaxy.  With a total gas mass
of $\sim 4 \times 10^6 M_\odot$ (Lis \& Goldsmith 1989)
concentrated within a region of size $\sim 30$~pc, Sgr B2
exhibits the largest submillimeter continuum flux of any astronomical
source outside the solar system.

The molecular inventory of the gas in Sgr B2 has been 
extensively studied by means of ground-based radio observations
(e.g.\ Cummins, Thaddeus \& Linke 1986; Turner 1989, 1991; 
Sutton et al.\ 1991; Nummelin et al.\ 1998, 2000).  Such 
observations have been supplemented by
far-infrared and submillimeter observations carried out
from airplane altitude or from space.  In particular,
observations carried out over the past decade with
the {\it Kuiper Airborne Observatory} (KAO; see Zmuidzinas et al.\ 1995), 
the {\it Infrared Space Observatory} (ISO; see Cernicharo et al. 1997)
and the {\it Submillimeter Wave Astronomy Satellite} (SWAS; see Neufeld
et al.\ 2000) have facilitated the observation of water vapor 
in Sgr B2. Water is of particular interest because of its
central role in the chemistry of interstellar oxygen and
its potential importance as a coolant of molecular clouds
(see, for example, Neufeld, Lepp \& Melnick 1995; Melnick et
al.\ 2000; Bergin et al.\ 2000).

In a earlier study of Sgr B2, 
we used SWAS to obtain the
first high-spectral-resolution ($\Delta v \sim 1 \kms$)
observations of H$_2^{16}$O
toward that source.  In a previous paper (Neufeld et al.\ 2000; hereafter N00), 
we reported the detection of water absorption over a wide range of 
LSR velocities that can be attributed to water vapor
in foreground clouds that are not associated with Sgr B2.
We have since extended these observations to map (incompletely) 
a region of size $26^\prime \times 19^\prime$ around Sgr B2. In the present
paper, we describe these new mapping observations and discuss
the abundance of water vapor in the vicinity of Sgr B2 itself.

\section{Observations}

Our mapping observations of Sagittarius B2 were carried out during 
the period 1999 March 20 through 2000 September 16.  They yielded 
incomplete coverage of a region extending $9.6^\prime$ North,
South and East and $16.0^\prime$ West of Sagittarius B2.
In Table 1 we list the on-source integration times for
each of 19 positions observed, together with the
offsets relative to the core of Sgr B2 at
position $\rm \alpha = 17^h\, 47^m \, 19.^s7$,
$\delta= -28^\circ 23^\prime 07^{\prime\prime}$ (J2000).  
At 550 GHz the SWAS beam size is $3.3^\prime \times 4.5^\prime$
and the main beam efficiency is 0.9 (Melnick et al.\ 2000).
All the data were acquired
in standard nodded observations (Melnick et al.\ 2000) and 
were reduced using the standard SWAS pipeline.  The
reference (nod) position was located at $\rm \alpha = 17^h\, 54^m \, 12.^s2$,
$\delta= -29^\circ 42^\prime 19^{\prime\prime}$ (J2000)
in a region of devoid of strong CO emission\footnote{At this reference position, 
the antenna temperature for $^{12}$CO J=1--0 was less than 0.5 K at 
all velocities}. 
The total on-source integration time for these mapping observations
was 90.4 hours.

Because SWAS employs double-sideband receivers, the 
$^{13}$CO~$J=5-4$ transition at 550.927 GHz and the
H$_2$O $1_{10} - 1_{01}$ transition at 556.936~GHz are
observed simultaneously in the lower and upper sidebands
of one receiver (Receiver 2).  For the mapping observations,
the SWAS local oscillator (LO) setting was chosen 
to separate the $^{13}$CO emission from Sgr B2 (at
$v_{\rm LSR} \sim 65 \, \rm km \, s^{-1}$) from
any H$_2$O emission or absorption at LSR velocities 
$v_{\rm LSR} \ge 0$.  For the (0,0) position, several
additional LO settings had previously been 
used (see N00) to allow the region
$-145 {\rm \, km \, s^{-1}}  \le v_{\rm LSR} \le 0 {\rm \, km \, s^{-1}}$ 
to be covered as well without contamination by $^{13}$CO emission from the
lower sideband.

\section{Results}

In Figure 1, we show the nineteen H$_2$O spectra obtained
from our mapping observations described above.  Each panel
shows a spectrum covering the range 
$-145 {\rm \, km \, s^{-1}}  \le v_{\rm LSR} \le 145 {\rm \, km \, s^{-1}}$.
The vertical dotted line denotes $v_{\rm LSR} \sim 65 \, \rm km \, s^{-1}$,
the systemic velocity of Sgr B2.  The upper horizontal dotted lines indicate
the continuum flux level, determined from the spectral regions 
$v_{\rm LSR} = -145$ to $-125\, \rm km \, s^{-1}$ and 
125 to 145$\, \rm km \, s^{-1}$ that are devoid of absorbing gas, and 
the lower horizontal dotted lines indicate
the zero flux level, estimated by the method described in \S 4.1 below. 
On each panel a pair of numbers, X/Y,
describe respectively the continuum antenna temperature 
(which corresponds to the separation
between the two dotted horizontal lines), and the
antenna temperature (in K) at the top of the panel.
The former is also given in Table 1.

The feature at $v_{\rm LSR} \sim - 90$ to $- 50 \,\rm km \, s^{-1}$ is a $^{13}$CO 
emission feature detected in the lower sideband.  In Figure 2, we
show the Receiver 2 data reduced as appropriate for the $^{13}$CO
line in the lower sideband;
a strong feature at $v_{\rm LSR} \sim 65 \,\rm km \, s^{-1}$ is observed
at several positions close to Sgr B2.  We note that the spectra
in Figure 2 are not perfect mirror images of those shown 
in Figure 1; in particular, the $^{13}$CO  features are significantly 
narrower in the spectra that were reduced as appropriate for the
lower sideband (Figure 2).  The reason for this behavior is that
the SWAS LO setting is fixed in the rest frame of the spacecraft,
which has a motion relative to the local standard of rest
(LSR) that varies quite rapidly.   In particular, the spacecraft 
orbits the Earth every 97 minutes with an orbital speed 
of $7\, \rm km \, s^{-1}$ (and -- of course --
the Earth also orbits the Sun.)  In the standard SWAS pipeline, the 
necessary Doppler corrections are applied to every 2 seconds' worth 
of data, and the sign of the correction is {\it opposite} for the lower and
upper sidebands.  Thus any spectral line originating in the lower sideband will be
smeared out as it appears in Figure 1, while any line from the upper sideband
will appear broadened in Figure 2.

Although not the primary emphasis of this paper, for completeness
we present additional spectra obtained for the  
$\rm ^3P_1 - ^3P_0$ fine structure
line of atomic carbon.  These spectra, shown in Figure 3, are 
obtained from Receiver 1 (upper sideband) on the SWAS instrument, which performs
simultaneous observations of the CI $\rm ^3P_1 - ^3P_0$ line
in a linear polarization orthogonal to that observed with Receiver 2.
There is no evidence for emission or absorption by the 487.249 $3_1 - 3_2$
transition of O$_2$, to which Receiver 1 is sensitive in its lower 
sideband.

\section{Discussion}

Even a cursory inspection of Figure 1 reveals three significant
features in the water spectra:  (1) extended absorption, observed
at almost every one of the 19 positions, covering the range $-100 \,{\rm km \, s^{-1}}
\simlt v_{\rm LSR} \simlt 0$; (2) extended emission, again observed at
almost every position, covering the range $80 
\simlt v_{\rm LSR} \simlt 120\,{\rm km \, s^{-1}}$; 
(3) less-extended absorption, only observed 
unequivocally at the $(0.0^\prime,0.0^\prime)$, $(0.0,+3.2^\prime)$, $(0.0,-3.2^\prime)$,
$(-3.2^\prime,0.0)$ and  $(-3.2^\prime,-3.2^\prime)$ 
positions, centered at 
v$_{\rm LSR} \sim 65 \,{\rm km \, s^{-1}}$, the systemic velocity of the 
Sgr B2 cloud.

\subsection{Extended absorption by water vapor at $\bf v_{\bf LSR} \sim 0 \, km\, s^{-1}$}

Broad absorption by water vapor over the entire velocity range
$v_{\rm LSR}$ = --100 to $0\, \rm km \, s^{-1}$ has been
inferred previously from SWAS observations of H$_2^{16}$O and H$_2^{18}$O 
toward Sgr B2 (N00; see also Figure 7), the H$_2^{18}$O  spectrum 
indicating that the peak optical depth occurs at $v_{\rm LSR} \sim 0$.
The absorption in this velocity range can be attributed to material lying at
a wide range of locations along the line-of-sight between the Solar circle 
and Sgr B2.  

The mapping observations presented in Figure 1 show clear evidence
for absorption at $v_{\rm LSR} \sim 0\, \rm km \, s^{-1}$ 
at 18 of the 19 positions
observed, implying that both the continuum emission and 
intervening material have a large angular extent.  The angular
extent of the absorbing water vapor at $v_{\rm LSR} \sim 0$ is 
considerably larger than that of the water vapor at $v_{\rm LSR} \sim 65 \kms$,
the systemic velocity of Sgr B2. 
The large spatial extent of the water absorption at 
$v_{\rm LSR} \sim 0\, \rm km \, s^{-1}$ is further demonstrated by 
the SWAS spectrum of water vapor toward Sgr A, shown in Figure 4, 
which clearly exhibits strong absorption in this velocity range.
Located at position $\rm \alpha = 17^h\, 45^m \, 39.^s7$,
$\delta= -29^\circ 00^\prime 30^{\prime\prime}$ (J2000), SgrA
is at offset (--21.8$^\prime$, --37.4$^\prime$) 
relative to Sgr B2.  Thus the
SWAS observations of the Galactic Center region show water
absorption at $v_{\rm LSR} \sim 0\, \rm km \, s^{-1}$ 
covering a region of longest angular dimension 50 arcmin, 
corresponding to a linear scale of 125 pc for an assumed distance
to the Galactic Center of 8.5 Kpc.  Of course, if the absorbing
material lies significantly closer to us than the Galactic Center,
then its linear extent could be smaller than 125 pc.

The presence of strong absorption at $v_{\rm LSR} \sim 0$ allows
us to estimate the continuum flux level, a quantity that is 
not reliably determined by the SWAS instrument. 
In particular, the zero flux level in Figure 1 and Table 1
is established on the assumption that 
the spectral region $v_{\rm LSR} =$ --5 to 5$\, \rm km \, s^{-1}$ is
subject to complete absorption by water vapor in foreground material
along the line-of-sight to Sgr B2.  This assumption is motivated by
the facts that (1) absorption at $v_{\rm LSR} \sim 0 \, \rm km \, s^{-1}$ 
is observed at almost every position and is well separated
from contamination from $^{13}$CO emission in the lower sideband; 
and (2) the H$_2^{18}$O spectrum of the $1_{10}-1_{01}$ line
at the (0.0$^\prime$,0.0$^\prime$) 
position (see N00; also Figure 7) implies an optical depth 
$\sim 1$ at $v_{\rm LSR} \sim 0$, corresponding to an H$_2^{16}$O $1_{10}-1_{01}$
optical depth of several hundred.  
If the $v_{\rm LSR} = $ --5 to 5$\, \rm km \, s^{-1}$
spectral region is not subject to complete absorption, then 
the continuum antenna temperatures quoted are lower limits.

The continuum antenna temperatures obtained by this method
can be compared with continuum maps presented by Dowell et al.\ (1999), who
observed Sgr B2 with 
the SHARC instrument at the Caltech Submillimeter Observatory.
In Table 1, we show the 350 $\mu$m continuum fluxes (in kJy per SWAS beam) --
obtained by convolving the SHARC data with the SWAS beam profile\footnote{approximated as
a Gaussian circular beam of diameter 4 arcmin (FWHM)} -- along
with the SWAS-determined 550~GHz ( 540 $\mu$m) continuum fluxes.  
The SHARC-measured 350 $\mu$m continuum 
fluxes should be regarded as lower limits because the relatively
small offset of only 2 arcmin
for the SHARC reference beam 
(Dowell, private communication) can lead to highly extended emission 
being ``chopped out"\footnote{A further uncertainty in the comparison of
the SWAS and SHARC continuum measurements
results from the different assumptions used in calibrating these measurements.
The SHARC data calibration effectively assumes that the beam is coupled to 
a point source, while the SWAS data is calibrated 
assuming that the source fills the main beam of the antenna.}. 
Toward the central 9 positions [i.e. the (0,0) position and 8 positions 
closest to it], the inferred 540 $\mu$m / 350 $\mu$m continuum flux
ratios all lie in the range 0.17 to 0.42.   
We have compared these values with 
those expected for optically-thin
dust with an opacity that varies as $\lambda^{-2}$.  
The ratios observed toward the central nine positions
are all consistent with dust temperatures of 15~K and above; the larger values 
of the 540 $\mu$m / 350 $\mu$m flux ratio observed at greater distances from
the cloud center suggest that the dust temperature is smaller
in the outer parts of the Sgr B2 cloud.

\subsection{Extended emission from water vapor in the 180-pc molecular ring  
($\bf v_{\bf LSR} \sim 80 - 120 \, \bf km \, s^{-1}$)}

Extended water emission in the $v_{\rm LSR} \sim 80 - 120 \, \rm km \, s^{-1}$
velocity range is detected from 18 of the 19 positions we observed in the vicinity
of Sgr B2, with the integrated antenna temperatures given in Table 2. 
Emission in this velocity range is also evident in the spectrum of Sgr A (Figure 4), 
although it is blended with a strong emission feature centered at 
$v_{\rm LSR} \sim 65 \, \rm km \, s^{-1}$.
In Figure 5, we show the average spectrum for those 14 positions near Sgr B2
that are devoid of apparent absorption at the Sgr B2 systemic velocity.
The solid line is Gaussian fit to the average emission feature:  
its velocity centroid is $\rm 93 \, km \,s^{-1}$, the line width is 
$\rm 43 \, km \,s^{-1}$, the peak antenna temperature is 0.075~K,
and the velocity-integrated antenna temperature is $3.4 \rm \, K \, km \,s^{-1}$.

This extended emission feature is kinematically similar to 
a highly extended emission component detected in $^{13}$CO maps
of the Galactic Center region (Bally et al.\ 1987; 
Lis \& Goldsmith 1989, hereafter LG89).  It arises from the far side
of a molecular ring of angular diameter $\sim$ 2 degrees
-- corresponding to a projected radius of $\sim 180$~pc -- 
the near 
side of which shows $^{13}$CO emission and H$_2$O absorption (N00)
at negative LSR velocities.  The kinematic properties of
this 180-pc molecular ring yield a parallelogram
in the $l-v$ plane; they have been attributed either to an
explosion-driven expansion (Scoville 1972; Kaifu et al.\ 1972)
or -- more recently -- to the gravitational effects of a
Galactic bar potential (Binney et al.\ 1991).

In Table 2, we also show the H$_2$O $1_{10} -  1_{01}$ / 
$^{13}$CO $J=1-0$ line ratio for each of the SWAS observed 
positions.  The $^{13}$CO $J=1-0$ spectra were obtained by
convolving the map of Bally et al.\ (1987) with the SWAS beam
profile, approximated as
a Gaussian circular beam of diameter 4 arcmin (FWHM).
For the 14 positions that are devoid of apparent
absorption at the Sgr B2 systemic velocity, the line ratios
apply to the velocity range $v_{\rm LSR} = 80 - 120 \, \rm 
km \, s^{-1}$.  For the other 5 observed positions, the
line ratios apply only to emission redward of the average
H$_2$O line centroid, i.e.\ to the velocity range 
$v_{\rm LSR} = 93 - 120 \, \rm km \, s^{-1}.$  The reason
for restricting the velocity range for these positions
is to avoid any effects of the Sgr B2 water absorption line
upon the line ratios derived for the extended emission 
feature.

With the exception of a single position at which 
the H$_2$O $1_{10} -  1_{01}$ line was not detected, the 
H$_2$O $1_{10} -  1_{01}$ / $^{13}$CO $J=1-0$ line ratio
lies within a factor 2 of 0.065.  In order to estimate the
water abundances implied by the observed H$_2$O $1_{10} -  1_{01}$ 
/ $^{13}$CO $J=1-0$ line ratios, we have made use of a
statistical equilibrium calculation for the line strengths.
We adopted the rate coefficients of Flower \& Launay (1985) for collisional
excitation of CO by H$_2$, and those of Phillips, Maluendes \& Green (1996)
for that of H$_2$O by H$_2$.\footnote{For both molecules, rate
coefficients are separately available for collisions with ortho- 
and para-H$_2$.  In the case of CO, the calculated rate coefficients
are very similar for collisions with ortho- and para-H$_2$, and thus the 
results of the statistical equilibrium calculation show little sensitivity
to the assumed ortho-to-ratio ratio (OPR) for H$_2$.  For
H$_2$O, by contrast, the rate coefficients for excitation
by ortho-H$_2$ are typically an order of magnitude larger than
those for excitation by para-H$_2$; 
the derived water
abundances therefore depend
upon the assumed H$_2$ OPR.  Following Snell et al.\ (2000), 
we adopted
the H$_2$ OPR expected in local thermodynamic equilibrium at
the gas temperature.}  
Our excitation model indicates
that the $^{13}$CO $J=1-0$ 
line is optically thin and that the H$_2$O $1_{10} -  1_{01}$ 
is effectively thin\footnote{i.e. we assume that each collisional excitation of
the H$_2$O $1_{10} -  1_{01}$ transition is followed by the
emission of a line photon, even though the line is optically thick.}.  
Thus the inferred H$_2$O to $^{13}$CO
abundance ratio is proportional to the H$_2$O $1_{10} -  1_{01}$ 
/ $^{13}$CO $J=1-0$ line ratio.  

The inferred H$_2$O to $^{13}$CO
abundance ratio also depends upon the conditions of temperature and
density assumed for the emitting region.  Based upon multitransition
observations of CO using the AST/RO telescope, Kim et al.\ (2002)
have argued that the physical conditions within a few hundred
parsec of the Galactic Center are remarkably constant -- except
within the Sgr A and Sgr B regions themselves -- with the gas temperature
everywhere in the range 30 to 45 K and the H$_2$
density\footnote{Towards 4 positions [(--3.2$^\prime$,--3.2$^\prime$), 
(0.0$^\prime$,--9.6$^\prime$), (0.0$^\prime$,--6.4$^\prime$), and 
(0.0$^\prime$,--3.2$^\prime$)],
SWAS clearly detects $^{13}$CO $J= 5-4$ emission in the  80 -- 120 km s$^{-1}$ range.
Assuming a gas temperature of $\sim 35$ K and a plausible range
of $^{13}$CO column density (10$^{15} - 10^{17}$ cm$^{-2}$), we find that the  
$^{13}$CO $J=5-4$/$J=1-0$ ratio is consistent with a H$_2$ density of 2000 -- 
4000 cm$^{-3}$.   This is in good agreement with the range estimated by 
Kim et al (2002).} in the range 2500 to 4000 $\rm cm^{-3}$.
Given these ranges of temperature and density, we
find that the ortho-H$_2$O/$^{13}$CO  abundance ratio is a
factor of 4 -- 27 times as large as the H$_2$O $1_{10} -  1_{01}$ 
/ $^{13}$CO $J=1-0$ line ratio.  The typical observed H$_2$O $1_{10} -  1_{01}$ 
/ $^{13}$CO $J=1-0$ line ratio of 0.065 therefore implies an ortho-H$_2$O/$^{13}$CO  
abundance ratio of 0.26 to 1.8.  Assuming a $^{12}$CO abundance of 10$^{-4}$
relative to H$_2$, a $^{12}$CO/$^{13}$CO abundance ratio of 30 for Galactic
Center material (Langer \& Penzias 1990), and an ortho-to-para ratio of
3 for water vapor, we obtain a value of 1.2 -- 8 $\times 10^{-6}$ for
the typical total water abundance relative to H$_2$ within the
``180-pc" molecular ring.  

These water abundances inferred for the
low density gas responsible for the $v_{\rm LSR} 
\sim 80 - 120 \, \rm km \, s^{-1}$ extended water emission
are two to four orders of magnitude 
larger than those measured within cold dense molecular clouds 
(e.g. Snell et al.\ 2000).  Thus the
results obtained here from observations of water {\it emission}
confirm our previous conclusion (N00) -- based upon {\it absorption}
line observations -- that the water abundance is typically
larger in low-density interstellar gas.  
The most plausible explanation lies in the density
dependence of the water vapor abundance, which is illustrated in
Figure 6.  This figure shows model predictions from the gas-grain model
presented by Bergin et al. (2000) for three different molecular
hydrogen densities.  For lower densities the longer depletion timescales
allow water vapor to exist in higher abundance.
In dense regions, the abundance of water vapor is suppressed more rapidly
by the depletion of oxygen nuclei
onto icy grain mantles (Bergin et al.\ 2000).

\subsection{Water abundance in Sagittarius B2}

In a previous paper (N00), we published the spectra of 
H$_2$O and H$_2^{18}$O $1_{01} - 1_{01}$ transitions
obtained toward the $(0,0)$ position 
and shown here in Figure 7.  Now that the general properties (i.e.\
velocity centroid and line width) of
the extended 80 -- 120~$\rm \, km \, s^{-1}$ emission have
been elucidated, we return to a consideration of the H$_2$O and H$_2^{18}$O
$1_{01} - 1_{01}$
absorption features at $v_{\rm LSR} \sim 65 \, \rm km \, s^{-1}$. 

We have attempted to fit the SWAS-observed absorption at 
the $(0,0)$ position with the use of two models.  In the first
model, we follow Zmuidzinas et al.\ (1995, hereafter Z95; 
see also Neufeld et al.\ 1997)
in assuming that the absorbing water vapor is 
mixed with the continuum-emitting dust in the core of 
Sgr B2 itself; whereas in the second model, we assume the water
vapor to lie in separate intervening clouds -- located
close to Sgr B2 -- that partially 
cover the continuum emission (see Ceccarelli et al.\ 2002).

\subsubsection{Water vapor in the core of Sgr B2}

In modeling the core of Sgr B2, we made use of a radiative transfer code
very kindly made available by J.~Zmuidzinas.  
This code yields a solution to the equations of statistical equilibrium for the water
level populations -- as a function of position within a spherically-symmetric, dusty,
molecular cloud core -- using the method of Accelerated Lambda Iteration (ALI) 
to obtain a converged, self-consistent solution to the level populations and the
radiation field. Ray tracing is then used to determine the 
spectrum observed by a telescope with a Gaussian beam profile of specified
projected diameter.  
The density in the
core of Sgr B2 was assumed to follow the profile of LG89, Model C, and 
the temperature of the gas and dust
to follow the profiles assumed by Z95.  The total continuum opacity was 
adjusted to match the observed (optically-thin)
continuum emission at 550~GHz, with
$\tau$ assumed to vary with wavelength as $\lambda ^{-1.5}$.
We adopted the values given by Phillips, Maluendes \& Green (1996)
for the rate coefficients for excitation of H$_2$O in collisions
with H$_2$, assuming an local thermodynamic equilibrium (LTE)
ortho-to-para ratio for H$_2$.  The excitation model included the
lowest 8 states of ortho-H$_2$O.

In fitting the SWAS-observed H$_2^{18}$O line, we varied
three parameters: the microturbulent line width in the cloud core, 
the H$_2^{18}$O abundance, and the velocity centroid relative to
the LSR.  All three were assumed to be constant throughout the
cloud core.  The blue curve in Figure 7 (lower panel) shows
the best fit to the H$_2^{18}$O absorption at $v_{\rm LSR} \sim 65 \rm\, 
km \, s^{-1}$ obtained by this method.  The best-fit microturbulent
line width is $20 \kms$, the ortho-H$_2^{18}$O abundance is $6.4 \times
10^{-10}$ relative to H$_2$, and the LSR velocity centroid is $64 \kms$.
Assuming an H$_2^{16}$O/H$_2^{18}$O abundance ratio of 250 
(N00; Langer \& Penzias 1990),
and a water ortho-to-para ratio of 3, we derive a total
H$_2^{16}$O abundance of $2.1 \times 10^{-7}$ relative to H$_2$.  
This derived abundance
is in acceptable agreement with the value of $3.3 \times 10^{-7}$
derived previously (Z95) from KAO
observations of H$_2^{18}$O and reported\footnote{In order to
fit the strengths of other, higher-excitation, H$_2^{18}$O lines 
detected toward this source, Neufeld et al.\ (1997) assumed that a 
higher water abundance of $5 \times 10^{-6}$ was present in the
inner parts of the Sgr B2 core, where the dust temperature was
sufficient ($\simgt 90$~K) to cause the vaporization of water ice
from grain mantles.  Because the absorption in the $1_{10}-1_{01}$
is completely dominated by material cooler than 90~K, the assumed
presence or absence of enhanced water abundances in the region warmer 
than 90~K has no effect upon the fitting procedure adopted here.} by N00.

In the upper panel of Figure 7, the blue curve shows the fit to
the H$_2^{16}$O absorption at $v_{\rm LSR} \sim 65 \kms$, for the
water abundance, line width and centroid velocity given above.
One additional free parameter is introduced here: the peak
antenna temperature (taken as 0.18~K in this fit)
of an emission component with the line
centroid and line width derived in \S 3.4 above for the 
``180-pc molecular ring".  
The fit to the H$_2^{16}$O absorption (upper panel, Figure 7, blue curve)
is clearly deficient in three 
regards: (1) the predicted absorption line is too broad; (2) the 
predicted absorption line is too deep,
and (3) the predicted emission peaks at $v_{\rm LSR} = 33$ and $95 \kms$ are
not observed.

The first of these deficiencies is a direct consequence of the simplified
velocity structure assumed in this model.  Observations of ammonia absorption
towards Sgr B2 (e.g. H\"uttemeister et al.\ 1995, hereafter H95)
have revealed the presence of three distinct absorption
components in the $v_{\rm LSR} = 60$ to $70\kms$ interval (see discussion
in the next subsection).  Although
none of these components shows a line width greater than $\sim 12 \kms$ (FWHM),
collectively they blend to impart an overall line width $\sim 20 \kms$ (FWHM)
to optically-thin absorption lines.  For the optically-thick H$_2^{16}$O line,
however, the observed line width does not increase as rapidly as it would for a
single Gaussian component. 

Taken at face value, the second deficiency -- the fact that the
observed H$_2$O line is less deep than the model predicts -- would imply
that the H$_2^{16}$O abundance is much less than assumed in the model.  
In Figure 8, middle panel, we show the line-center flux-to-continuum-flux 
ratio -- as predicted by the Zmuidzinas radiative transfer code -- as a function of 
the assumed water abundance.  
The observed  ratio $\sim 0.25$ would imply
an H$_2^{16}$O abundance of only $\sim 2 \times 10^{-9}$ and a very small H$_2^{16}$O/H$_2^{18}$O
ratio of $\sim 3$.  We are aware of no chemical fractionation mechanism 
capable of yielding such a small H$_2^{16}$O/H$_2^{18}$O ratio.  A far more likely
explanation of the anomalous depth of the H$_2^{16}$O absorption feature is
that some fraction ($\sim 25\%$) of the continuum emission detected in
the SWAS beam originates in material with a very small water abundance
(and which is not covered by any intervening cloud containing water).
The likely presence of continuum emission that is not covered by a 
significant water column density
has implications for the H$_2^{18}$O abundance obtained above.  For the water-covered
component, the effective line-center to continuum ratio decreases from 
0.49 to $(0.49 - 0.25)/ 0.75 = 0.32$, and the implied ortho-H$_2^{18}$O abundance 
increases by a factor of almost $2$ from $6.4 \times 10^{-10}$ to $1.1 \times 10^{-9}$.
Given the same ortho-to-para and H$_2^{16}$O/H$_2^{18}$O ratios assumed previously, 
this corresponds to a total H$_2^{16}$O abundance of $3.7 \times 10^{-7}$.  Therefore, 
we  conservatively
adopt the range $2 - 4 \times 10^{-7}$ as the H$_2^{16}$O abundance
implied by our model of the Sgr B2 core.

The third deficiency of the simple model we considered is its prediction of
narrow emission peaks at $v_{\rm LSR} = 35$ and $95 \kms$; this again is
inconsistent with observed spectrum.  A ``double-horned" structure
is predicted because the $1_{10} - 1_{01}$ resonance-line photons that are 
excited collisionally within the cloud core are scattered significantly
in frequency before they can escape the cloud.   The top and bottom
panels in Figure 8 show the predicted line-to-continuum ratios in these peaks and
the predicted frequency offsets from line center (in velocity units).

The observed absence of the blue-shifted horn is naturally explained
by the known presence of foreground absorbing material at 
$v_{\rm LSR} \sim 33\kms$.  The absence of a red-shifted horn
could easily be explained if the velocity dispersion increases
slightly with increasing distance from the core center; such
an increase would lead to the red-horn photons being spread over 
an spatially-extended region, leading to a decrease in the
observed surface brightness relative to what is predicted by our 
constant velocity-dispersion model (see discussion in N00).

\subsubsection{Water vapor in an intervening cloud}

As an alternative means of interpreting the H$_2$O and
H$_2^{18}$O absorption towards the $(0,0)$ position in Sgr B2, we
have fitted the spectrum with a series of absorption features
that are assumed to originate in foreground molecular gas
outside the core of Sgr B2.  In this picture -- favored by 
H95 and by Ceccarelli et al.\
(2002; hereafter C02) for the origin of the NH$_3$ absorption 
observed toward Sgr B2 -- the absorption originates in warm,
low-density gas located outside the central core of Sgr B2 itself.
According to C02, ISO
observations of far-infrared NH$_3$ absorption lines imply
a gas temperature $\sim 700$~K,
while the absence of observable far-IR CO emission places
an upper limit $n({\rm H}_2) \simlt  3 \times 10^3 \rm \,cm^{-3}$ on the gas 
density; in the best-fit model of C02,
the warm absorbing gas is located at a distance $\sim 1.15$~pc
from the center of the Sgr B2 core.

In fitting the SWAS-observed H$_2$O and H$_2^{18}$O spectra in 
this picture, we adopted the absorption line parameters derived
by H95 from high resolution observations of NH$_3$ absorption
lines.  Those observations were carried out separately 
toward two condensations within the Sgr B2 cloud: Sgr B2(N) and Sgr B2(M).
These condensations, which are separated by $\sim 0.8^\prime$, 
are not resolved individually by the $3.3^\prime \times 4.5^\prime$ SWAS beam.  
H95 observed absorption components
at $v_{\rm LSR} = 60.4$ and $69.0\kms$ toward Sgr B2(M) but not Sgr B2(N),
and absorption components
at $v_{\rm LSR} = 66.7$ and $81.5\kms$ toward Sgr B2(N) but not Sgr B2(M).
The line widths and NH$_3$ column densities 
inferred by H95 for these absorption components are listed in Table 3.

We regarded the velocity centroids and line widths for these
four components to be fixed in our fitting of the SWAS-observed 
H$_2$O and H$_2^{18}$O spectra.  We took as adjustable parameters
the following: (1) the fraction, $f_A$, of the continuum emission covered
by the 60.4 and $69.0\kms$ clouds\footnote{The $60.4$ and $69.0\kms$ clouds 
were paired in this manner -- with a
common assumed covering factor -- and the 66.7 and $81.5\kms$ clouds likewise
paired, because the H95 observations suggest that the former pair
cover Sgr B2(M) but not Sgr B2(N), and the latter pair cover Sgr B2(N) but 
not Sgr B2(M)}; (2) the fraction $f_B$, of the 
continuum emission covered by the 66.7 and $81.5\kms$ clouds; (3) the
strength of the 80 -- 120$\kms$ emission feature at the $(0,0)$ position in
Sgr B2 (with an assumed line centroid and width as derived from the 
average spectrum shown in Figure 5 and discussed in \S 4.2); (4) the
H$_2^{18}$O line center optical depths in each of the four absorption
components.  
Following C02, the absorption was considered to occur in a ``foreground
screen". 

As in \S4.3.1, we assumed an H$_2^{16}$O/H$_2^{18}$O 
abundance ratio of 250, and a water ortho-to-para ratio of 3.
In the upper panel of Figure 7, the red curve shows our best fit to
the H$_2^{16}$O absorption using this model.  Because the
H$_2^{16}$O lines have optical depths of several hundred
and therefore lie on the flattest part of the curve-of-growth,
the H$_2^{16}$O spectrum is almost entirely insensitive
to the assumed optical depth; it therefore provides a relatively
unambiguous determination of the covering factors $f_A$ and $f_B$.
The red curve represents our best fit to the H$_2^{16}$O spectrum,
obtained for $f_A= 0.6$, $f_B=0.2$, and with a 80 -- $120 \kms$ emission
feature of peak antenna temperature 0.2~K.

Once the covering factors have been determined from the H$_2^{16}$O
spectrum, the line-center optical depths can be determined
from the H$_2^{18}$O spectrum.  The red curve in the lower
panel of Figure 7 represents our best fit to the data, 
obtained with line-center optical depth of 1.3, 1.0, 0.8, and 1.5
respectively for the $v_{\rm LSR} = 60.4$, 66.7, 69.0, and $81.5\kms$
features.  

ISO observations of the $2_{21} - 1_{10}$ 
transition of H$_2^{16}$O at 108.073 $\mu$m
yield a further observational constraint upon the material responsible
for the water absorption.  A Long-Wavelength 
Spectrometer Fabry-Perot spectrum of the source 
(Goicoechea \& Cernicharo 2003) yields
an absorption line equivalent width $\sim 20\kms$.  Using a simple excitation
model to estimate the fractional populations of the H$_2^{16}$O 
rotational states, we find the observed 108 $\mu$m
equivalent width to be consistent with
the best-fit optical depths and covering factors 
derived from the SWAS observations (Table 3).  Unfortunately,
the 108 $\mu$m line lies on the flat part of the curve-of-growth,
so the observed equivalent width does not place strong constraints
upon the physical conditions in the absorbing material.

The absorbing column densities implied by 
the H$_2^{18}$O line-center depths are inversely proportional to the
difference $f_{01} - f_{10}$, where $f_{01}$ is the fraction
of ortho-water molecules in the 1$_{01}$ rotational state 
and $f_{10}$ is the fraction in 1$_{10}$.  Thus the minimum
column density of H$_2^{18}$O is obtained for the assumption
$f_{01}=1$, $f_{10} = 0$; these minimum values are tabulated
in Table 3.  The beam-averaged, minimum total column density -- weighted
by the continuum antenna temperature -- is $N({\rm H_2^{18}O}) = 1.0 \times 
10^{14} \rm \, cm^{-2}$, a value in excellent agreement with the
value of $1.1 \times 10^{14} \rm \, cm^{-2}$ derived previously
for the minimum H$_2^{18}$O column density from KAO observations by
Z95.  The actual column density is larger by a factor 
$(f_{01} - f_{10})^{-1}$. 

We have used a statistical equilibrium model for the
H$_2^{18}$O level populations
to estimate the correction
factor $(f_{01} - f_{10})^{-1}$.  
In Figure 9, we plot the 
values of $f_{01}$, $f_{10}$ and their difference $f_{01} - f_{10}$
as a function of the radiation field.  Here we follow C02 in
assuming that the far-infrared continuum radiation originates in
a region of radius $R_{\rm FIR}$ = 0.8 pc, and determine the H$_2^{18}$O level 
populations as a function of the ``dilution factor", $W$.  
The spectral shape of the far-infrared continuum 
source was obtained from Pipeline-10 archived ISO LWS grating
spectra and re-processed using the ISO Spectroscopic Analysis Package
version 2.1.\footnote{The ISO Spectroscopic Analysis Package is a joint
development by the LWS and SWS Instrument Teams and Data Centers.
Contributing institutes are CESR, IAS, IPAC, MPE, RAL and SRON.}
The dilution factor is defined such that the solid angle subtended by the 
far-infrared continuum source is $4\pi W$ at the location of the
absorbing water vapor.  It is related to the
distance, $R$, from the center of Sgr B2 (top horizontal axis in Figure 9)
by the expression $W=  \half [1 - (1 - R_{\rm FIR}^2/R^2)^{1/2}]$.
Solid curves in Figure 9 show results for the temperature and density favored
by C02 for the absorbing material: an H$_2$ density of $3 \times 10^3 \rm
\, cm^{-3}$ and a gas temperature of 700~K.   
Dotted curves 
show the results obtained for the same assumed temperature but
with the density increased by a factor of 10. 

Two features of Figure 9 are noteworthy.  First, the level populations
are not greatly affected by a factor 10 increase in the assumed 
density.  This behavior is explained by the fact that the gas density
is very much smaller than the ``critical density" for the 
$1_{10}-1_{01}$ transition; the excitation of this transition 
is therefore controlled
by the radiation field.  Second, the quantity $f_{01} - f_{10}$
is close to unity unless the absorbing material is located very 
close to the center of the Sgr B2 core.  For the distance of 1.15~pc
favored by C02, corresponding to $W=0.15$, 
the value of $f_{01} - f_{10}$ is 0.8 and the correction factor
$(f_{01} - f_{10})^{-1}$ is only 1.25.  Significantly smaller
values of the distance of the absorbing material from the cloud core
are ruled out by the requirement that the absorbing water vapor must
lie in front of most of the emitting dust.   We conservatively adopt
an upper limit of 1.5 for the correction factor
$(f_{01} - f_{10})^{-1}$.  

Using the analysis method of \S4.3.2, where the water
absorption is assumed to arise in a foreground cloud
of warm, low-density material, our estimate of the
beam-averaged
H$_2^{18}$O column density lies in the range 
$1.0 -  1.5 \times 10^{14} \, \rm cm^{-2}$.
\footnote{We note that
this value is
substantially smaller than an estimate of $10^{15} \, \rm cm^{-2}$
given by C02 for $N({\rm H_2^{18}O})$ but agrees with
that obtained by Comito et al.\ (2003).}  Given an assumed abundance
ratio of 250 for H$_2^{16}$O/H$_2^{18}$O, the H$_2^{18}$O column density
we derived implies an H$_2^{16}$O column density of 
$2.5 -  4 \times 10^{16} \rm \, cm^{-2}$. 
The total H$_2$ column density of the warm absorbing gas layer
is not well constrained by observations; if we adopt the
estimate of $10^{22} \rm \, cm^{-2}$ given by Flower et al.\ (1995),
then the water abundance is a few $\times 10^{-6}$.

As noted by Flower et al.\
(1995) -- and contrary to the assertion
of C02 that there is no discrepancy -- 
the observed water column density lies an order 
of magnitude below the
predictions of shock models for the warm, absorbing gas layer
in front of Sgr B2.  
Recent observations of high OH abundances
in Sgr B2 (Goicoechea \& Cernicharo 2002) may provide an explanation
of the discrepancy between theory and observation.  As suggested
by Goicoechea \& Cernicharo, if the shocked gas is irradiated by
strong ultraviolet radiation from a population of hot stars, 
then photodissociation will reduce the H$_2$O abundance and increase 
the OH abundance.  Further chemical
modeling is needed, however, to determine whether this explanation is
either quantitatively plausible
or consistent with the large observed abundance of NH$_3$ in the
absorbing gas.

\subsubsection{Summary: water vapor in the cold core or in 
a warm intervening cloud?}

In \S4.3.1 and \S 4.3.2 above, we have used two alternative
methods to analyse the SWAS observations of H$_2^{16}$O
and H$_2^{18}$O at the $(0,0)$ position in Sgr B2.
Although the first method -- in which the water vapor and dust
are assumed to be well mixed in the cloud core -- yields a
fit to the data which is inferior to that of the second method --
in which the water vapor is assumed to lie in a foreground gas
cloud -- the poorer fit can be ascribed to the fact that
the assumed velocity structure was far simpler in the
first analysis method.  On the basis of the SWAS data alone, therefore,
we do not find compelling evidence to favor one picture
over the other.\footnote{We note, however, that Comito et al.\
(2003) have carried out an analysis
which combines observational data from SWAS with ground-based
observations of HDO and para-H$_2^{18}$O.  They conclude
that a significant fraction of the water absorption must
take place within warm foreground material.}

In principle, both the cold core and a warm intervening
cloud might contribute to the observed water absorption.
Insofar as the water column densities involved add linearly,
we may conclude that
$$\biggl( {N_{\rm fg}(\rm{H_2O}) \over {\rm 2.5\,to\,4 \times 10^{16}  cm^{-2}}} \biggr)
+ \biggl( {x_{\rm c}(\rm{H_2O}) \over \rm 2 - 4 \times 10^{-7}} \biggr) \sim 1 \eqno(1) $$
where $N_{\rm fg}(\rm{H_2O})$ is the water column density in the warm foreground
material and $x_{\rm c}(\rm{H_2O})$ is the water abundance relative to H$_2$ in
the core.  Clearly, we may regard $4 \times 10^{16} \, \rm cm^{-2}$
and $4 \times 10^{-7}$ as conservative upper limits on $N_{\rm fg}(\rm{H_2O})$
and $x_{\rm c}(\rm{H_2O})$.\footnote{Similar remarks apply to the
far-IR hydrogen fluoride absorption observed by Neufeld et al.\ (1997) toward 
Sgr B2.  Assuming the HF absorption to occur entirely in the core of Sgr B2, 
Neufeld et al.\ derived a low HF abundance that implied a substantial
depletion factor $\sim 50$ for fluorine in the dense core.  C02 pointed
out (correctly) that a considerably larger HF abundance would be implied
if the absorption were assumed to take place in intervening material at
a distance $\sim 1.15$~pc from the core center, but argued (erroneously)
that this would weaken the argument for a high fluorine depletion
in the core of Sgr B2.
Clearly, if part or all of the observed HF absorption occurs in foreground
material {\it outside} the core of Sgr B2, then the HF abundance {\it within} the
core would have to be {\it even less} than that inferred by Neufeld et al. (1997).
This would merely {\it strengthen} the argument for a high fluorine depletion 
in the core of Sgr B2.}

\begin{acknowledgments}

We gratefully acknowledge the support 
of the SWAS Science Team at Smithsonian Astrophysical Observatory,
and particularly Zhong Wang, Frank Bensch and John Howe.
We thank Jonas Zmuidzinas for making
available his radiative transfer code.
We thank Claudia Comito for her comments on an earlier version of the
manuscript and for communicating results to us prior to their publication.
We are grateful to the anonymous referee for several helpful suggestions.
This work was supported NASA's SWAS contract NAS5-30702. 
The National Astronomy and Ionosphere Center is operated by Cornell University
under a Cooperative Agreement with the National Science Foundation. 
\end{acknowledgments}

\clearpage

\begin{figure}
\vspace{7.5in}
\special{vscale=85.0 hscale=85.0 hoffset=-15.0 voffset=-80.0 psfile=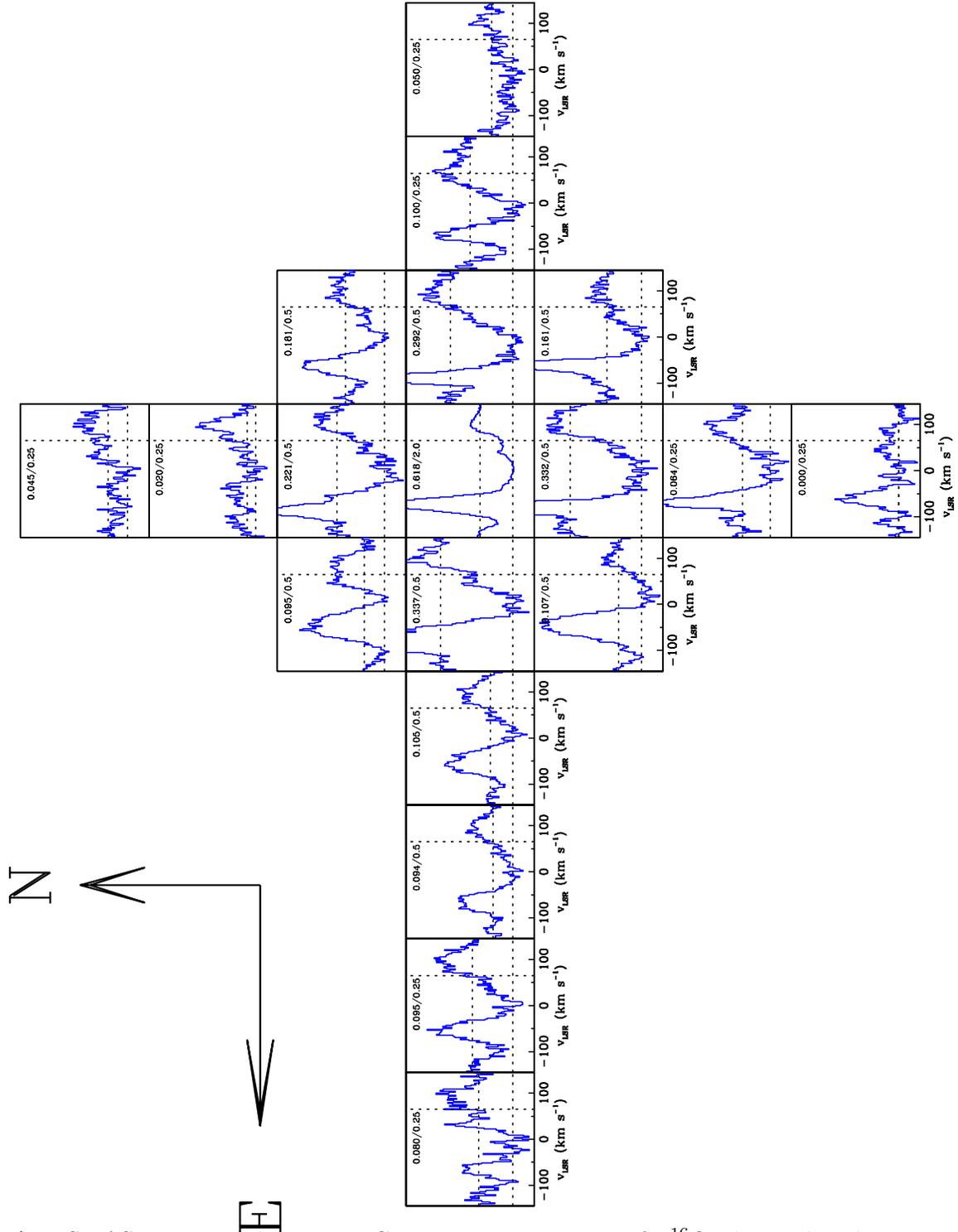}
\caption{\small SWAS spectra of the 556.936 GHz $1_{10}- 1_{01}$
transition of H$_2^{16}$O, observed in the vicinity of Sgr B2.
Each panel covers the LSR velocity range $-145$ to $145 \kms$. 
Adjacent spectra are offset by $3.2^\prime$.
Horizontal dotted lines indicate the
the continuum flux level and 
the zero flux level.  The vertical dotted lines indicate a LSR velocity
of 65$\,\rm km \, s^{-1}$.
On each panel, a pair of numbers, X/Y, 
describe, in order, the continuum antenna temperature 
and the antenna temperature (in K) at the top of the panel.  
The feature at $v_{\rm LSR} \sim - 90$ 
to $- 50 \,\rm km \, s^{-1}$ is a $^{13}$CO 
emission feature detected in the lower sideband.
The (0,0) position is in the middle of the central 
group of nine spectra.}
\end{figure}

\clearpage

\begin{figure}
\vspace{7.5in}
\special{vscale=85.0 hscale=85.0 hoffset=-15.0 voffset=-80.0 psfile=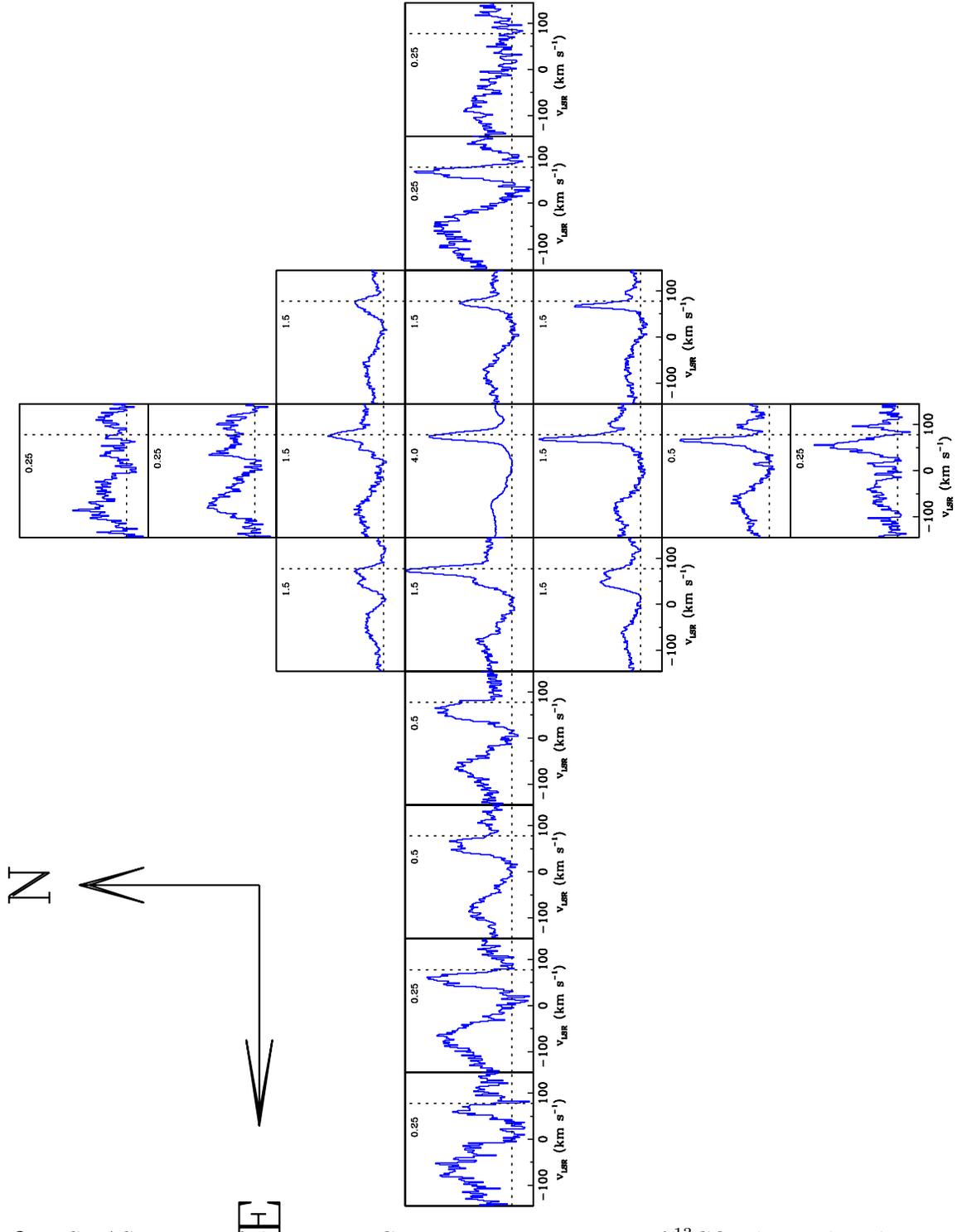}
\caption{\small SWAS spectra of the 550.927 GHz $J = 5 -4$
transition of $^{13}$CO, observed in the vicinity of Sgr B2.
Each panel covers the LSR velocity range $-145$ to $145 \kms$. 
The vertical dotted lines indicate a LSR velocity
of 65$\,\rm km \, s^{-1}$.
Adjacent spectra are offset by $3.2^\prime$.
The number on each panel gives 
the antenna temperature (in K) at the top of the panel.
The feature at $v_{\rm LSR} \sim - 120$ 
to $- 80 \,\rm km \, s^{-1}$ is H$_2$O  
emission being detected in the upper sideband.
The (0,0) position is in the middle of the central 
group of nine spectra.}
\end{figure}

\clearpage

\begin{figure}
\vspace{7.5in}
\special{vscale=85.0 hscale=85.0 hoffset=-15.0 voffset=-80.0 psfile=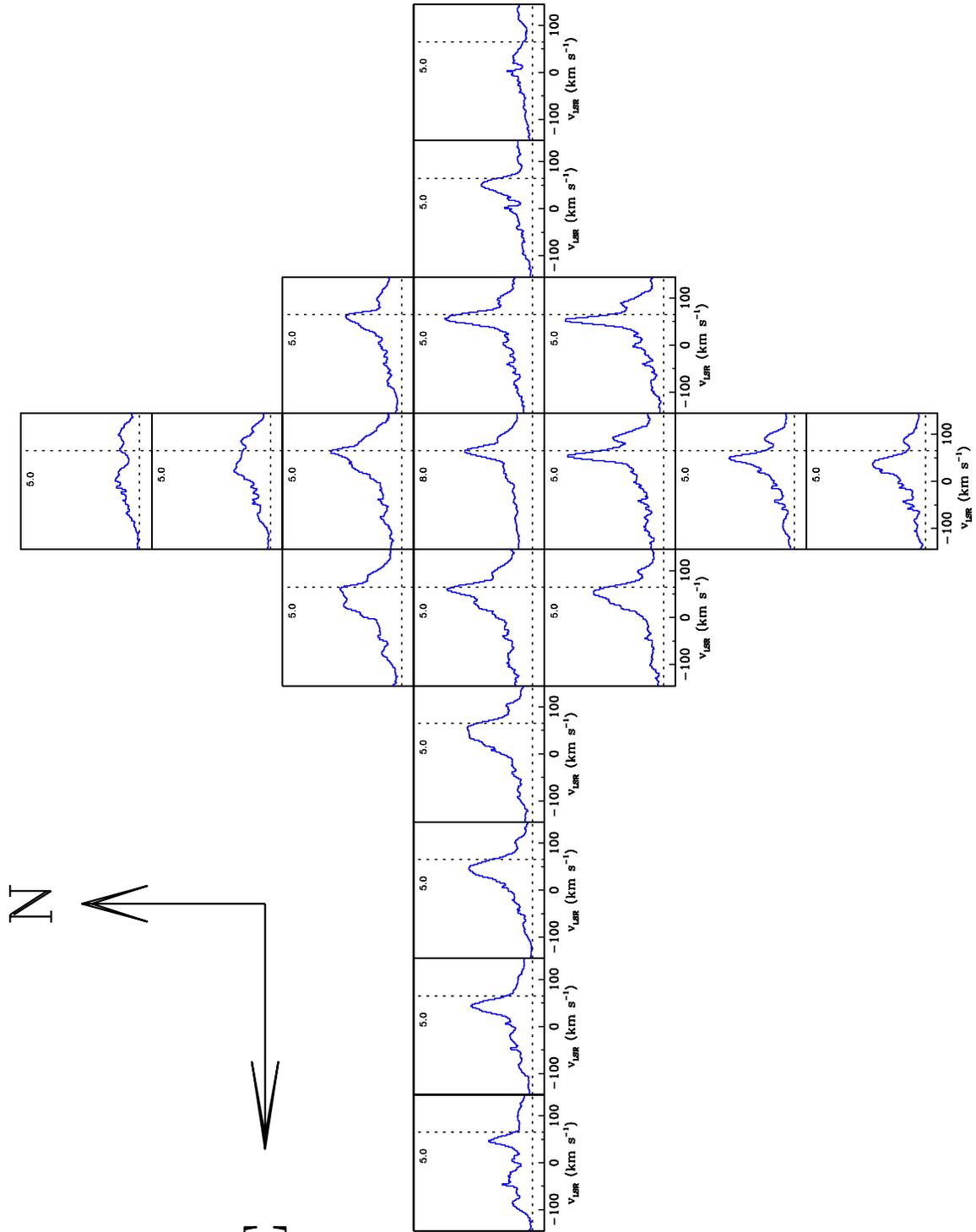}
\caption{\small Same as Figure 2, but for the $\rm ^3P_1-^3P_0$ fine structure
line of atomic carbon.}
\end{figure}
\clearpage

\begin{figure}
\vspace{4.5in}
\special{vscale=60.0 hscale=60.0 hoffset=0.0 voffset=40.0 angle=0 psfile=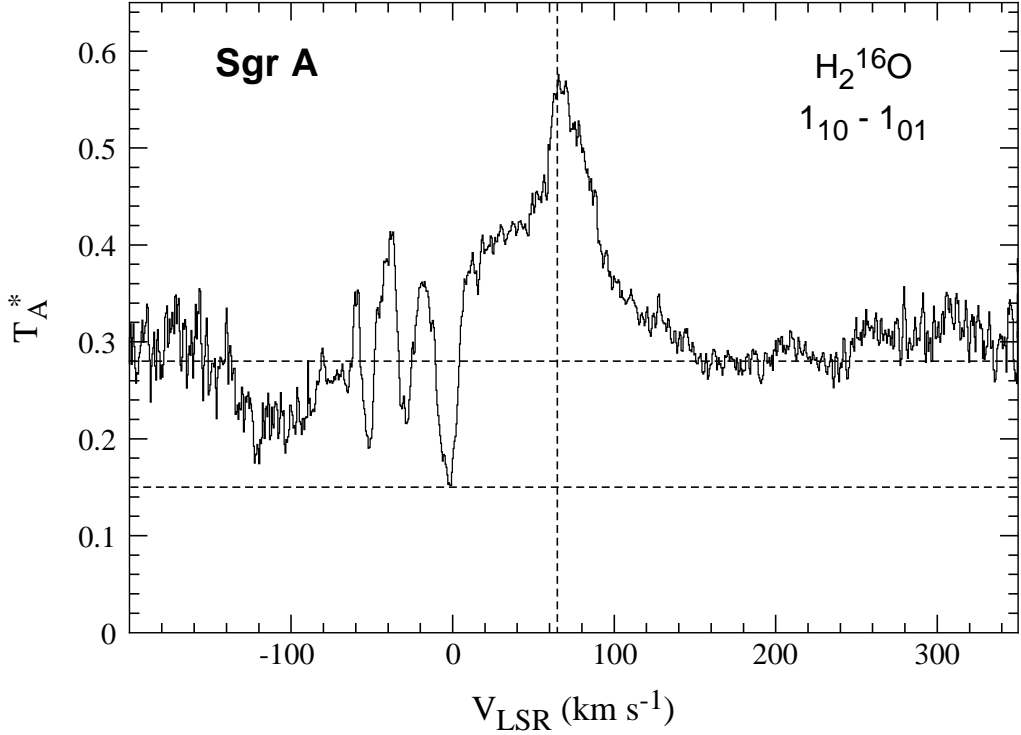}
\caption{
SWAS H$_2^{16}$O spectrum of Sgr A, with the beam centered
at $\rm \alpha = 17^h\, 45^m \, 39.^s7$,
$\delta= -29^\circ 00^\prime 30^{\prime\prime}$ (J2000), corresponding
to an offset (--21.8$^\prime$, --37.4$^\prime$) relative to Sgr B2. 
This spectrum was obtained in 454 hours of total
integration time at 11 separate local oscillator (LO) settings
during the months of July 1999, April 2000,  
May 2000, August 2000, and Sept. 2000.  The 
$^{13}$CO J=5--4 and H$_2^{16}$O $1_{10}-1_{01}$ lines 
were detected simultaneously in different sidebands of SWAS Receiver 2 
(see text).
The use of different LO frequencies 
allowed us to reconstruct the true $^{13}$CO J=5--4 line profile free
from contamination by water emission or absorption.
This profile was then subtracted  
from the H$_2^{16}$O spectrum  
(with proper account taken of the varying spacecraft velocity and different
LO settings used), yielding 
a H$_2^{16}$O spectrum free of $^{13}$CO emission which has a
considerably larger bandwidth ($\sim 500$ km s$^{-1}$) than a typical
SWAS spectrum ($\sim 210$ km s$^{-1}$).
The rms noise in the spectrum varies
from 0.015 K for $v_{\rm LSR}$ in the range --60 to 300 km s$^{-1}$ 
to 0.05 K for velocities outside that range.}
\end{figure}
\clearpage

\begin{figure}
\vspace{4.5in}
\special{vscale=60.0 hscale=60.0 hoffset=0.0 voffset=380.0 angle=-90 psfile=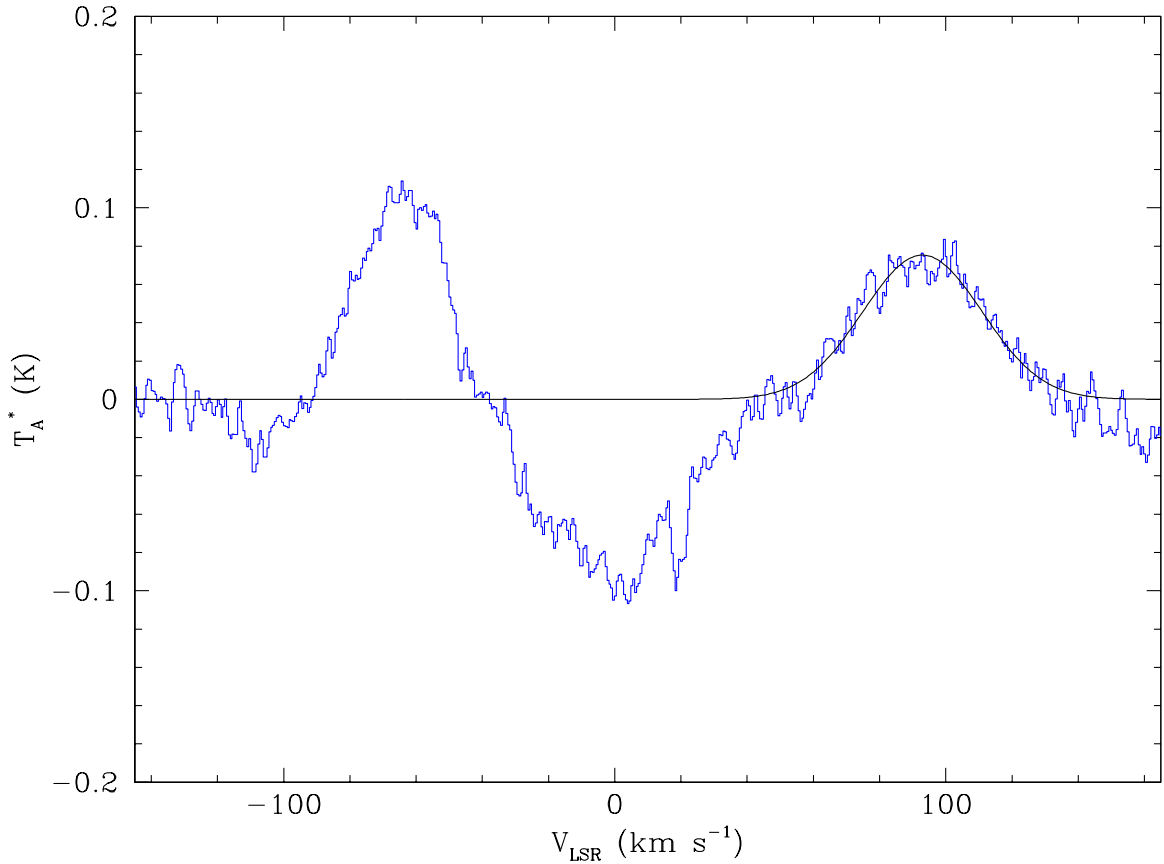}
\caption{Average H$_2^{16}$O $1_{10} - 1_{01}$ spectrum for those 14 positions
that are devoid of absorption at the Sgr B2 systemic velocity.
The solid line is a Gaussian fit to the average emission feature with
the following parameters:  
velocity centroid $\rm 93 \, km \,s^{-1}$, line width  
$\rm 43 \, km \,s^{-1}$ (FWHM), peak antenna temperature 0.075~K,
and velocity-integrated antenna temperature $3.4 \rm \, K \, km \,s^{-1}$.
The observed emission is believed to orignate in the far side of the ``180-pc" molecular ring.}
\end{figure}  

\begin{figure}
\epsscale{0.9}
\plotone{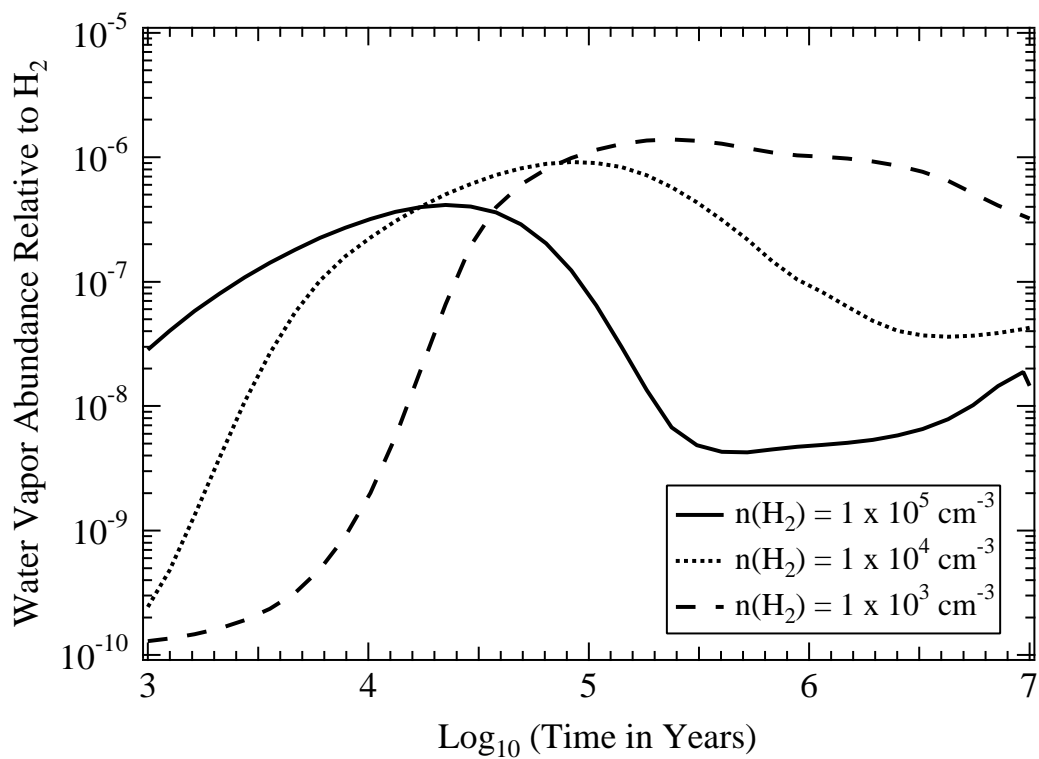}
\caption{
Water vapor abundance as a function of time for three different H$_2$ 
densities.  Model results are predicted using the gas-grain model with
limited surface chemistry presented by Bergin et al.\ (2000).  The gas is
assumed to be completely shielded from external ultraviolet radiation 
and to have a temperature of 10~K.
}
\end{figure}

\begin{figure}
\epsscale{0.7}
\plotone{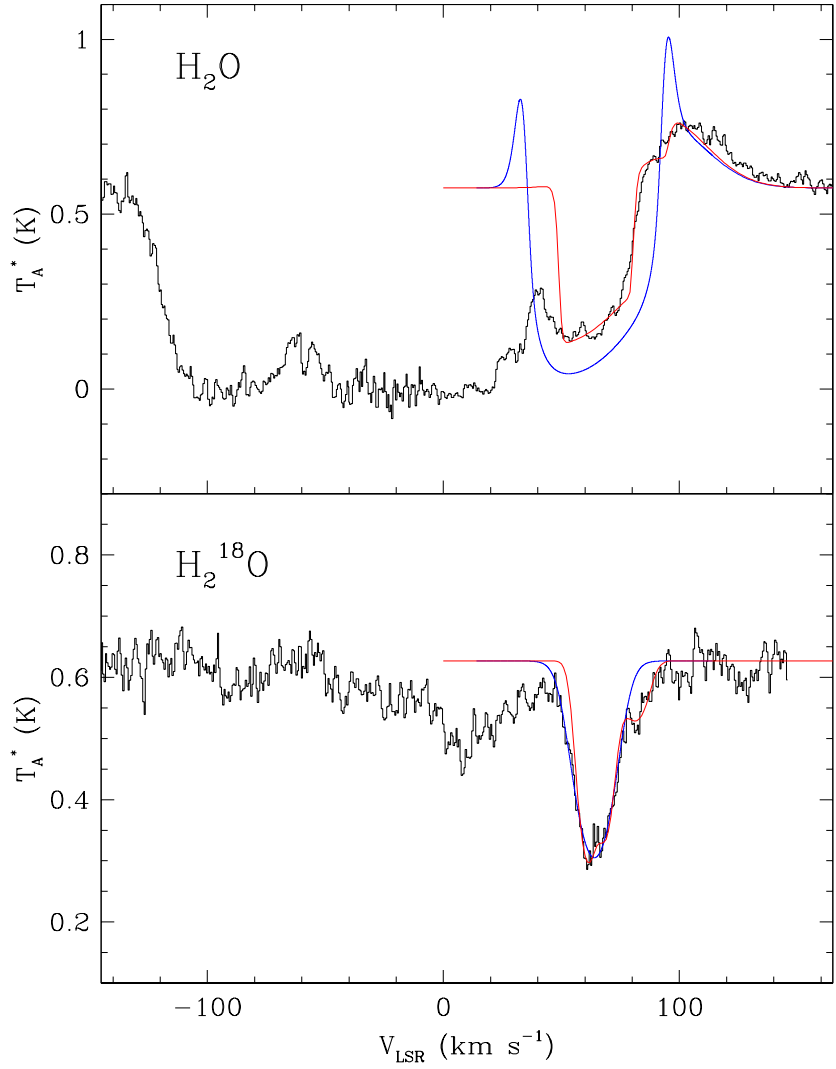}
\caption{SWAS H$_2^{16}$O and H$_2^{18}$O $1_{10} - 1_{01}$ spectra
of the Sgr B2 $(0,0)$ position (from the paper of N00).  Blue
and red curves show the fits described in the text.}
\end{figure}

\begin{figure}

\vspace{4.5in}
\special{vscale=60.0 hscale=60.0 hoffset=50.0 voffset=00.0 angle=0 psfile=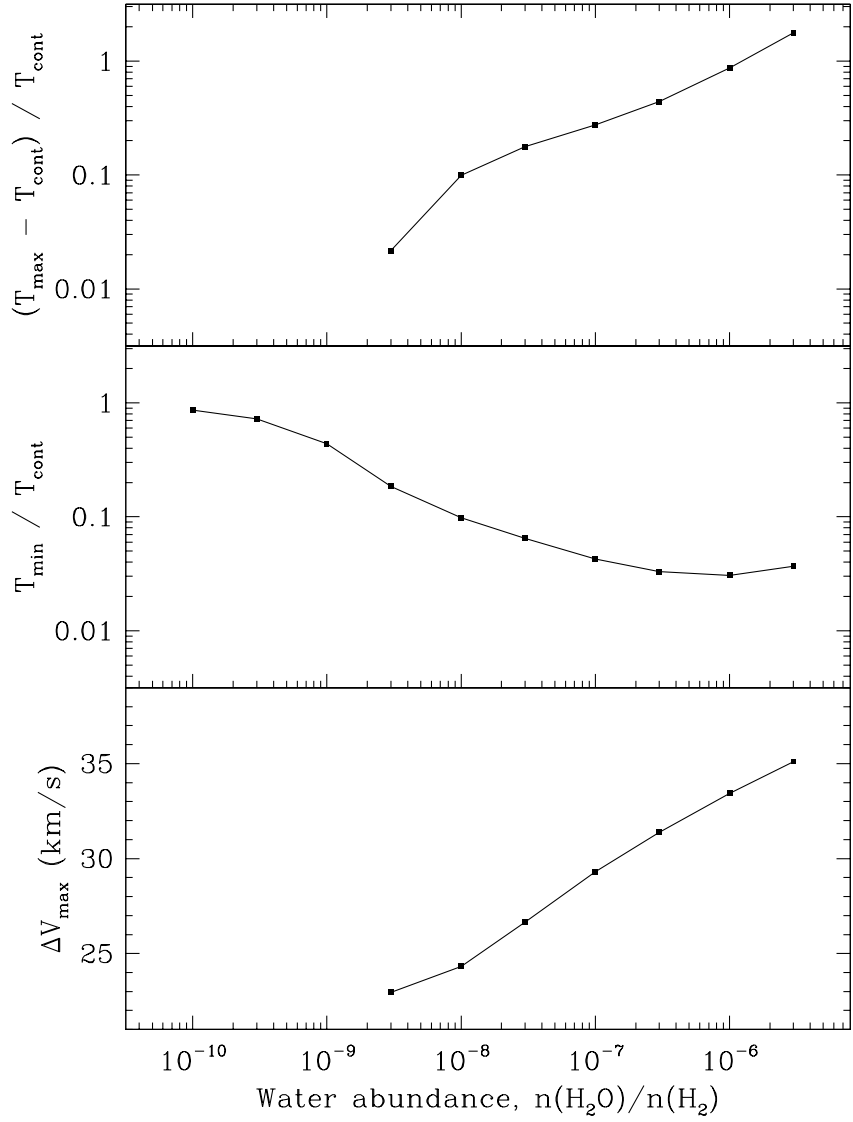}
\caption{Predictions of the Zmuidzinas radiative transfer code for Sgr B2.  
Top panel: peak line-to-continuum ratio for the double horns, as a function of
the total water abundance.  Middle panel: ratio of line-center flux to 
continuum flux, as a function of the total water abundance.  Lower panel: 
velocity separation ofthe double horns from line center.}
\end{figure}

\begin{figure}
\vspace{4.5in}
\special{vscale=70.0 hscale=70.0 hoffset=0.0 voffset=40.0 angle=0 psfile=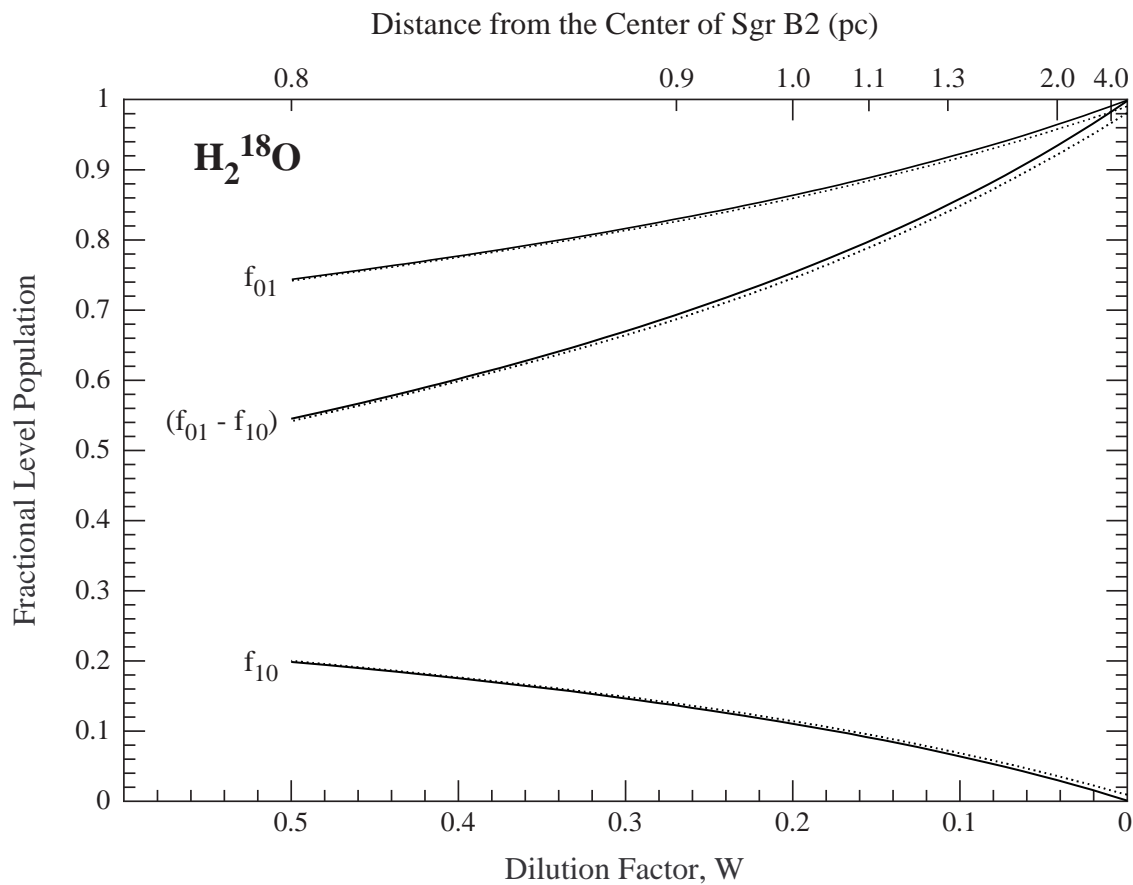}
\caption{Fractional level populations, $f_{01}$ and $f_{10}$,
and their difference $f_{01} - f_{10}$,
as a function of distance from the center of Sgr B2 (top horizontal axis;
the bottom horizontal axis shows the corresponding 
``dilution factor" for the radiation field [see text]).  
Solid curves
apply to an H$_2$ density of $3 \times 10^3$~cm$^{-3}$
and a gas temperature of 700~K, values favored by C02.  Dotted curves 
show the results obtained for the same assumed temperature but
with the density increased by a factor 10; the results are virtually
unchanged because the molecular excitation is controlled by the radiation
field, not collisional excitation.}
\end{figure}

\clearpage


\begin{tabular}{c c c c c} 
\multicolumn{5}{c}{TABLE 1}\\
\multicolumn{5}{c}{H$_2^{16}$O observations toward Sgr B2: inferred continuum fluxes}\\
\\
\hline 
\\
Offset$^a$&Time $^b$&Continuum $T_A^*$ $^c$& 550 GHz continuum& 350 $\rm \mu m$ continuum\\
&(min)&(mK)& flux$^d$ (kJy) & flux$^e$ (kJy)\\
\\
\hline 
\\
$(-9.6, +0.0)$& 394 &  50 & 0.52 & 0.06\\
$(-6.4, +0.0)$& 332 & 100 & 1.04 & 1.71\\
$(-3.2, -3.2)$& 157 & 161 & 1.67 & 7.49\\
$(-3.2, +0.0)$& 107 & 292 & 3.03 & 10.8\\
$(-3.2, +3.2)$& 263 & 181 & 1.88 & 4.90\\
$(+0.0, -9.6)$& 404 & $< 20$ & $<0.3$ & 1.77\\
$(+0.0, -6.4)$& 495 &  64 & 0.66 & 3.27\\
$(+0.0, -3.2)$&  99 & 332 & 3.45 & 12.7\\
$(+0.0, +0.0)$& 373 & 618 & 6.42 & 31.9\\
$(+0.0, +3.2)$& 100 & 221 & 2.29 & 11.0\\
$(+0.0, +6.4)$& 487 &  20 & 0.21 & 1.02\\
$(+0.0, +9.6)$& 391 &  45 & 0.47 & 0.03\\
$(+3.2, -3.2)$& 263 & 107 & 1.11 & 3.82\\
$(+3.2, +0.0)$& 102 & 337 & 3.50 & 8.32\\
$(+3.2, +3.2)$& 261 &  95 & 0.99 & 5.68\\
$(+6.4, +0.0)$& 202 & 105 & 1.09 & 1.78\\
$(+9.6, +0.0)$& 379 &  94 & 0.98 & 0.10\\
$(+12.8, +0.0)$& 394 &  95& 0.99 & $<$0.025 \\
$(+16.0, +0.0)$& 225 &  80& 0.83 & $<$0.025 \\

\\

\hline
\\

\multicolumn{5}{l}{$^a$ Offset ($\alpha\cos\delta, \delta$) in arcmin
relative to $\rm \alpha = 17^h\, 47^m \, 19.^s7$,
$\delta= -28^\circ 23^\prime 07^{\prime\prime}$ (J2000)}\\
\multicolumn{5}{l}{$^b$ On-source integration time}\\
\multicolumn{5}{l}{$^c$ Antenna temperature at 550 GHz}\\
\multicolumn{5}{l}{$^d$ Measured by SWAS }\\
\multicolumn{5}{l}{$^e$ Measured by SHARC (Dowell et al. 1999)}\\

\end{tabular}

\begin{tabular}{c c c} 
\multicolumn{3}{c}{TABLE 2}\\
\multicolumn{3}{c}{H$_2^{16}$O observations toward Sgr B2: emission line strengths}\\
\\
\hline 
\\
Offset$^a$&$\int_{80}^{120} (T_A^*-T_{AC}^*) dv$ $^b$ 
& H$_2$O/$^{13}$CO  \\
&(K~$\kms$) & line ratio$^c$ \\
\\
\hline 
\\
$(-9.6, +0.0)$& 0.74 & 0.074\\
$(-6.4, +0.0)$& 1.07 & 0.046\\
$(-3.2, -3.2)^d$& 1.66 & 0.035\\
$(-3.2, +0.0)^d$& 3.33 & 0.053\\
$(-3.2, +3.2)$& 1.50 & 0.046\\
$(+0.0, -9.6)$& $<$ 0.76$^e$ & $<$ 0.031$^e$\\
$(+0.0, -6.4)$& 2.15 & 0.043\\
$(+0.0, -3.2)^d$& 2.89 & 0.044\\
$(+0.0, +0.0)^d$& 4.18 & 0.103\\
$(+0.0, +3.2)^d$& 2.31 & 0.071\\
$(+0.0, +6.4)$& 3.04 & 0.060\\
$(+0.0, +9.6)$& 1.98 & 0.058\\
$(+3.2, -3.2)$& 1.49 & 0.036\\
$(+3.2, +0.0)$& 3.50 & 0.070\\
$(+3.2, +3.2)$& 3.53 & 0.071\\
$(+6.4, +0.0)$& 4.09 & 0.098\\
$(+9.6, +0.0)$& 3.34 & 0.125\\
$(+12.8, +0.0)$&2.43 & 0.101\\
$(+16.0, +0.0)$&2.16 & 0.089\\

\\

\hline
\\

\multicolumn{3}{l}{$^a$ Offset ($\alpha\cos\delta, \delta$) in arcmin
relative to $\rm \alpha = 17^h\, 47^m \, 19.^s7$,
$\delta= -28^\circ 23^\prime 07^{\prime\prime}$ (J2000)}\\
\multicolumn{3}{l}{$^b$ Continuum-subtracted $T_A^*$,
integrated over the interval $v_{\rm LSR} = 80$ to 120 $\kms$}\\
\multicolumn{3}{l}{$^c$ Ratio of continuum-subtracted $T_A^*$,
integrated over the interval $v_{\rm LSR} = 80$ to 120 }\\
\multicolumn{3}{l}{$\kms$, for the H$_2$O $1_{10}-1_{01}$ transition and the $^{13}$CO
$J=1-0$ transition}\\
\multicolumn{3}{l}{(Bally et al.\ 1987).  This ratio is computed using fluxes that
are uncorrected for }\\
\multicolumn{3}{l}{
antenna efficiency because the efficiencies of both
antennas are comparable }\\
\multicolumn{3}{l}{
(Bally, priv. comm.; Melnick et al 2000).}\\
\multicolumn{3}{l}{$^d$ Line ratio computed for the
smaller interval $v_{\rm LSR} = 93$ to 120 $\kms$ so as to}\\
\multicolumn{3}{l}{avoid the influence of water absorption near
the Sgr B2 systemic velocity (see text).}\\
\multicolumn{3}{l}{$^e$ 3 sigma upper limit}\\

\end{tabular}

\begin{tabular}{c c c c c } 
\multicolumn{5}{c}{TABLE 3}\\
\multicolumn{5}{c}{Water absorption line parameters}\\
\\
\hline 
\\
LSR velocity$^a$&Line width$^a$&Covering&$\tau$(H$_2^{18}$O) $^b$&Minimum\\
($\kms$)  &($\kms$ FWHM)  &factor&&$N({\rm H_2^{16}O})$ \\
\\
\hline 
\\
60.4 & 8.0  & 0.6 & 1.3 & $1.7 \times 10^{16}\,\rm cm^{-2}$ \\ 
66.7 & 12.1 & 0.2 & 1.0 & $2.0 \times 10^{16}\,\rm cm^{-2}$ \\
69.0 & 8.0  & 0.6 & 0.8 & $1.1 \times 10^{16}\,\rm cm^{-2}$ \\
81.5 & 9.8  & 0.2 & 1.5 & $2.5 \times 10^{16}\,\rm cm^{-2}$ \\

\\

\hline
\\

\multicolumn{5}{l}{$^a$ From H95}\\
\multicolumn{5}{l}{$^b$ H$_2^{18}$O optical depth at line center}\\
\end{tabular}

\end{document}